\documentclass[letterpaper, 10 pt, conference]{ieeeconf}  
\usepackage{graphicx}
\usepackage{xcolor}
\usepackage{subfig}
\usepackage{caption}
\usepackage{mathtools}
\usepackage{amsmath}
\usepackage{makecell}
\usepackage{hyperref}

\IEEEoverridecommandlockouts                              

\title{\LARGE \bf
Adipose Tissue Segmentation in Unlabeled Abdomen MRI using Cross Modality Domain Adaptation}

\author{Samira Masoudi$^{1,2}$, Syed M. Anwar $^{1}$, Stephanie A. Harmon $^{2,3}$, Peter L. Choyke $^{2}$,\\ Baris Turkbey$^{2}$, Ulas Bagci $^{1}$
\thanks{$^{1}$Masoudi, Anwar, and Bagci are with University of Central Florida, Orlando, FL 32816, USA  {\tt\small bagci@ucf.edu}}
\thanks{$^{2}$Masoudi, Harmon, Turkbey, and Choyke are with Molecular Imaging Program, National Cancer Institute, National Institutes of Health, Bethesda, MD 20814, USA {\tt\small ismail.turkbey@nih.gov }}
\thanks{$^{3}$Harmon is with Clinical Research Directorate, Frederick National Laboratory for Cancer Research, Frederick, MD 21701, USA  {\tt\small stephanie.harmon@nih.gov}}
\thanks{$^{*}$The implementation of this work can be found at: \url{https://github.com/SamMas-hub/Fat-Segmentation-in-MR-using-domain-adaptation}}
}

\begin{document}
\maketitle
\thispagestyle{empty}
\pagestyle{empty}

\begin{abstract}

Abdominal fat quantification is critical since multiple vital organs are located within this region. Although computed tomography (CT) is a highly sensitive modality to segment body fat, it involves ionizing radiations which makes magnetic resonance imaging (MRI) a preferable alternative for this purpose. Additionally, the superior soft tissue contrast in MRI could lead to more accurate results. Yet, it is highly labor intensive to segment fat in MRI scans. In this study, we propose an algorithm based on deep learning technique(s) to automatically quantify fat tissue from MR images through a cross modality adaptation. Our method does not require supervised labeling of MR scans, instead, we utilize a cycle generative adversarial network (C-GAN) to construct a pipeline that transforms the existing MR scans into their equivalent synthetic CT (s-CT) images where fat segmentation is relatively easier due to the descriptive nature of HU (hounsfield unit) in CT images. The fat segmentation results for MRI scans were evaluated by expert radiologist. Qualitative evaluation of our segmentation results shows average success score of $3.80/5$ and $4.54/5$ for visceral and subcutaneous fat segmentation in MR images$^{*}$. 
\end{abstract}

\section{INTRODUCTION}
\label{sec:intro}
Abdominal obesity is an increasingly prevalent condition (clinically considered as a disorder) among people of all ages including young adults and children. There is a significant body of evidence demonstrating a positive association between obesity indices and metabolic syndrome, higher risks of cancer, chronic inflammation, and diabetes as well as cardiovascular diseases. Visceral adipose tissue (VAT) and subcutaneous adipose tissue (SAT) have different roles in human metabolism, where VAT is known to be critically associated with major health risks~\cite{wajchenberg2000subcutaneous}. 

Therefore, quantifying the extents of adipose tissue in abdominal region in two forms of visceral and subcutaneous fat could help to better understand and evaluate a patient's condition~\cite{ibrahim2010subcutaneous}. The adipose tissue (SAT and VAT) can be identified using non-invasive imaging techniques such as computed tomography (CT), dual x-ray absorptiometry (DEXA), and magnetic resonance imaging (MRI). DEXA is the most widely used method for monitoring body fat with no direct information on anatomy nor separation between VAT and SAT~\cite{kelly2009dual}. Alternatively, by using either CT or MRI, the fat content as well as the underlying anatomy can be visualized. MRI is known to be a safer imaging modality since it does not use potentially harmful ionizing radiations, which is imminent during CT and DEXA acquisition. Although MRI is safer, anatomically informative, and highly accurate, it is difficult to segment fat in magnetic resonance (MR) images as compared to CT images. 

There are automated and semi-automated algorithms presented in literature to segment both VAT and SAT using different imaging modalities~\cite{hwang2018mr}, \cite{langner2019fully}. The contrast created by Hounsfield unit (HU) values in CT images is found to be appropriate and hence resulted in a widespread use of segmentation methods based on thresholding. However, the same methods cannot be applied to fat segmentation in MRI scans. Moreover, these methods produce approximate segmentation, which demands further manual inspection by clinical experts~\cite{langner2019fully}. Ideally, development of an automated segmentation algorithm using appropriate feature selection and machine learning techniques (such as deep learning) with high accuracy could save radiologists from a time-intensive workload.

Herein, we propose a fully automated algorithm for adipose tissue segmentation in MR images of the abdominal region. The training of such a network would require a large amount of manually labeled MR images. To alleviate this requirement, we propose to segment visceral and subcutaneous fat tissues in MR images using an unpaired set of labeled CT scans. For this purpose, we first employ the idea of cycle generative adversarial network (C-GAN) to construct a pipeline that transforms our target MRI scans into synthetic CT (s-CT) by generating 2D slices with similar (to a large extent) spatial distribution. Next, we use a U-Net, trained to segment VAT and SAT in s-CT and thus its reference MR scan. Results obtained using our proposed methodology in both cases were evaluated (using a set of defined rules) by expert radiologist. 
    \begin{figure*}[!ht]
      \centering
\includegraphics[width=1.\textwidth]{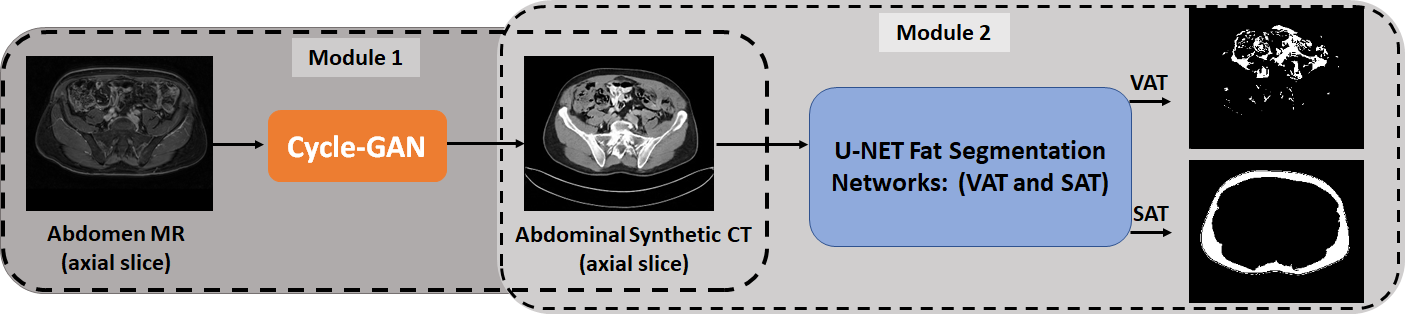}
\caption{\footnotesize{Our proposed solution for fat segmentation in unlabeled MR images using unpaired domain transformation and the U-Net architecture.}}
\label{fig:res2}
   \end{figure*}
 
Our algorithm has two major contributions: 1) we used an unpaired C-GAN architecture to generate abdominal s-CT images from input MR images, 2) we then utilized the underlying structural similarity (between input MR and s-CT images) to segment VAT and SAT in abdominal MRIs by using a U-Net model trained on the acquired CT (a-CT) data. Doing so, we obtained a significant performance in segmenting VAT and SAT in MR images without using labeled MR images for training. This work can significantly contribute towards solving problems (segmentation/detection) in certain radiology applications with scarce labeled data. 

\section{PROPOSED METHODOLOGY}
\label{sec:methods}

\begin{figure}[!b]

\includegraphics[width=.48\textwidth]{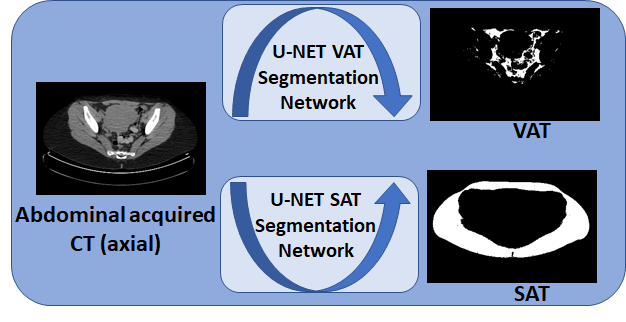}
\caption{\footnotesize{U-Net based fat segmentation networks (for VAT and SAT) trained using CT (acquired) scans and manually labeled ground truth.}}
\label{fig:res1}
  \end{figure}

We employed two main modules in our proposed methodology (Fig.~\ref{fig:res2}) for fat tissue segmentation in abdomen MRI. These two subsequent modules imply transformation and segmentation in 2D radiology scans. The first module is a C-GAN which we used to convert MR scans to their equivalent CT (s-CT). The second module (Fig.~\ref{fig:res1}) includes two separate U-Nets for independent segmentation of VAT and SAT in CT slices. We hypothesize that the segmentation labels for SAT and VAT from the s-CT could be transferred to the associated MR images. Since we trained our C-GAN with a loss function that maintains a spatial registration (between the input MRI and s-CT), while transforming the input MRI scan to s-CT, we believe that our hypothesis holds true. Our experimental results and expert radiologist validation added further credence to this hypothesis. The details on how these modules were trained and how the VAT/SAT segmentation was inferred for MRI scans are presented in the following subsections.

\subsection{How to Train?}

\subsubsection{Pre-processing}

The CT and MR scans were acquired from various centers, many patients at different depths, exposures, and physical conditions. Hence we utilized a prepossessing strategy to obtain meaningful intensity values among different MR images. To this end, we first used N-4 Bias correction algorithm to remove the low frequency non-uniformity or ``bias field" which is present in MR scans. We then performed a patient level standardization of intensity distribution, where intensity values in each image were transformed such that the resulting histogram was in harmony with a predefined standard distribution~\cite{tustison2010n4itk}, \cite{nyul1999standardizing}. These pre-processing steps significantly facilitated the learning process of our proposed C-GAN architecture. Similarly, for CT scans, we cropped the intensity values within a window (-700 HUs - 1300 HUs) belonging to the soft tissue (which is the focus of our segmentation). This was followed by intensity normalization in a fixed range among all patients. 

\subsubsection{MR to CT domain adaptation}
\label{subsubsec:met1_2}
Conventionally, learning an image-to-image mapping requires a large amount of paired data. In medical imaging domain, the availability of sufficient paired data is not always guaranteed. A C-GAN could overcome this limitation by allowing image translation without the requirement to have paired examples for training the model~\cite{zhu2017unpaired}. 

The C-GAN architecture was built upon the GAN model to enable dual-domain image translation among images from domain $A$ (MRI) and $B$ (CT). C-GAN uses two generators ($G_{A}$ generating $I_{BA}$ from $I_B$, and $G_{B}$ generating $I_{AB}$ from $I_A$) along with two discriminators ($D_{A}$ recognizing $I_A$ from $I_{BA}$ and $D_{B}$ recognizing $I_B$ from $I_{AB}$). We achieved an overreaching performance with this framework, by introducing a certain level of cycle consistency, to ensure successful image translation. Hence, we defined the loss function of our proposed generators using mean square error (MSE) given as, 
\begin{align}
     \label{Eq:Eq1} 
        \text{G}_\text{loss}= &\text{MSE}(\textbf{1},D_{B}(I_{AB}))+ \text{MSE}(\textbf{1},D_{A}(I_{BA}))\nonumber\\
        + \alpha&[\text{MSE}(I_{ABA},I_A)+ \text{MSE}(I_{BAB},I_B)]\nonumber\\
        + \beta &[\text{MSE}(I_{BA},I_{B})+\text{MSE}(I_{AB},I_A)],
\end{align}
\textcolor{black}{where $\alpha$ and $\beta$ are optimized to be $10.0$, and $2.0$ respectively. While $I_{AB}$, and $I_{BA}$ represents the synthetic CT and MR images generated from $I_A$ (MRI) and $I_B$ (CT) scans, \textcolor{black}{$I_{ABA}$ and $I_{BAB}$ refer to images generated after completion of the cyclic (C-GAN) process.}} Our proposed loss function was designed to ensure intensity transformation while maintaining spatial registration between input MRI and the generated s-CT. This spatial consistency between MRI and s-CT is crucial as we used this property to indirectly segment fat in MR images through their equivalent s-CT scans. For this module, we used fat saturated T1-weighted MR and CT scans obtained at the National Institute of Health (NIH) to train our network and employed it later to perform domain adaption among MR and CT images.  

\begin{table}[!t]

\caption{\footnotesize{The scoring rules for visual evaluation of synthetic CT (s-CT) scans generated using C-GAN.}}
    \label{tab:table0}
     \begin{center} 
    \begin{tabular}{|c|l|}
        \hline \textbf{Score} & \textbf{Description} \\
        \hline \hline
        1  & No correlation between MR and synthetic CT\\
        \hline
        2  & Few relations between MR and synthetic CT \\
        \hline
        3  & Some correlation between acquired CT and synthetic CT\\
        \hline
        4  & Comparable synthetic CT and acquired CT\\
        \hline
        5  & synthetic CT can substitute acquired CT\\
        \hline

    \end{tabular}
  \end{center}
\end{table}

\subsubsection{Fat segmentation in a-CT images}
\label{subsubsec:met1_1}
We trained a supervised machine learning based algorithm to segment fat tissue in abdominal CT scans. Image segmentation is an important problem in computer vision, and recently deep learning has demonstrated significant performance in this regard~\cite{liu2019auto}, \cite{chen2017deeplab}. One of the segmentation challenges is how to define the segment boundaries, which depends on the application and input data. Additionally, segmentation in medical image applications suffers from either limited labeled data or expensive annotations (in terms of time and labour). In recent years, the U-Net based architecture and its variants have shown to perform significantly well in medical image segmentation with limited training data~\cite{ronneberger2015u}, \cite{ibtehaz2020multiresunet}. 

Since VAT and SAT are localized at different body regions representing different properties, we employed a slightly different structure to optimally perform segmentation task in each case. We trained both of our deep learning based models by customizing the U-Net architecture; 3-layers and 5-layers for subcutaneous and visceral fat segmentation, respectively. While training from scratch, we included data augmentation in our training to overcome the limited availability of annotated data. 

We must note that it is relatively an easier task for radiologists to annotate fat (both VAT and SAT) in CT scans. That is why we used fat labels from CT scans to train our U-Net models, and transformed MRIs to s-CT for segmentation.  

\subsection{How to infer?}
\subsubsection{Step 1: Generating s-CT from MR scans}
\label{subsubsec:met2_1}
Although our C-GAN model was enforced to learn a reciprocal transformation between MR/CT slices, we are specifically interested in mapping MR images to CT. Hence, during the inference stage, we fed MR slices (which we want to segment) to the trained C-GAN to generate its equivalent s-CT.

\subsubsection{Step 2: Fat segmentation in MR images using s-CT}
\label{subsubsec:met2_2}
The s-CT images obtained in Step 1 were fed into the trained models (section~\ref{subsubsec:met1_1}) for segmenting visceral and subcutaneous fat. We hypothesize here that since the acquired MR and s-CT (generated using C-GAN) had an inherent spatial alignment, the segmented fat in s-CT (on a pixel level) can be directly assigned to the corresponding MR scan. Our results showed that this hypothesis holds true and we achieved a significant performance in segmenting fat in MR images. These finding were verified by expert observer adding credence to our hypothesis and results.

\begin{table}[!t]
\caption{\footnotesize{Terms used for visual evaluation of fat segmentation in MR slices based on false discovery rate (FDR) and false negative rate (FNR).}}
    \label{tab:table1}
    \begin{center} 
    \begin{tabular}{|c|l|c|}
        \hline
        \textbf{Score} & \textbf{Description} & \textbf{Ranges} \\
        \hline \hline
        1  & Complete failure & \thead{FDR$>70\%$ \& FNR$>80\%$} \\
        \hline
        2  & Considerable miss-predictions & \thead{$70\%>$FDR$>45\%$  \\ $80\%>$FNR$>50\%$} \\
        \hline
        3  &  Erroneous predictions & \thead{$45\%>$FDR$>30\%$  \\ $50\%>$FNR$>30\%$} \\
        \hline
        4  & Some  false predictions & \thead{$30\%>$FDR$>15\%$ \\ $30\%>$FNR$>20\%$} \\
        \hline
        5  & Very few miss-predictions &  \thead{FDR$<15\%$ \& FNR$<20\%$} \\
        \hline
    \end{tabular}
  \end{center}
\end{table}

\section{Results}

\subsection{Dataset}
In this study, two different datasets were anonymized and used for segmentation and transformation tasks: 1) we utilized CT and T1-weighted MR images of 34 patients obtained within 6 months of each other at NIH to train our C-GAN architecture. 
These scans had an average of 100 slices per patient (for both CT and MRI), among them we used $85\%$ for training, and $15\%$ to evaluate our proposed algorithm. 2) we used CT scans (from the abdomen region) of 131 patients obtained from multiple centers (Mayo clinic) with their respective visceral and subcutaneous fat labels for segmentation. We randomly split these patients into 90-10\%  as train-test data to train our segmentation networks based on U-Net. This data set was sufficient to train a network that could accurately segment subcutaneous fat. But the complexity of visceral fat segmentation required a larger number of training images for successful training. To compensate for the limited number of annotated CT slices, we employed data augmentation using image translation (width and height), reflection (horizontal flip), and zooming to increase (by 4 folds) our set of training images. 

\subsection{Evaluation metrics}
The performance was evaluated separately at each Step. We used dice score to report segmentation performance in a-CT images. During inference, we exploited the judgment of an expert radiologist to measure our success during Steps 1 (Section \ref{subsubsec:met2_1}) and 2 (Section \ref{subsubsec:met2_2}) of our proposed method. Our expert radiologist followed a set of defined scoring rules to evaluate the results of our proposed  MR to s-CT transformation design (Table \ref{tab:table0}). Similarly, Table~\ref{tab:table1} represents the respective definitions in terms of false discovery rate (FDR) and false negative rate (FNR) which are used to evaluate fat segmentation in MR scans. It must be noted that the ``correlation" in Table~\ref{tab:table0} was measured in terms of both visual structural and textural similarities. 
 
\subsection{Quantitative and qualitative results}
\begin{figure}[!t]
   \subfloat[\qquad\qquad\qquad\qquad(b)\qquad\qquad\qquad\qquad(c)]{\includegraphics[width=0.48\textwidth]{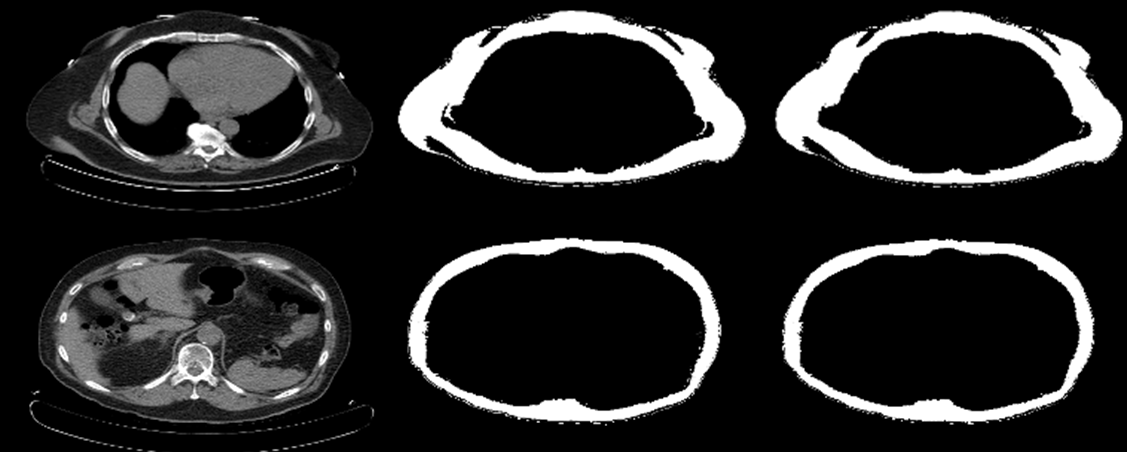}}
    \caption{\footnotesize{Examples of subcutaneous fat segmentation: (a) the acquired CT, (b) ground truth, and (c) segmented SAT using U-Net based architecture.}}
   \label{fig:fig_Unet_CT_Resutls}
\end{figure}

\begin{figure}
   \subfloat[\qquad\qquad\qquad\qquad(b)\qquad\qquad\qquad\qquad(c)]{\includegraphics[width=0.48\textwidth]{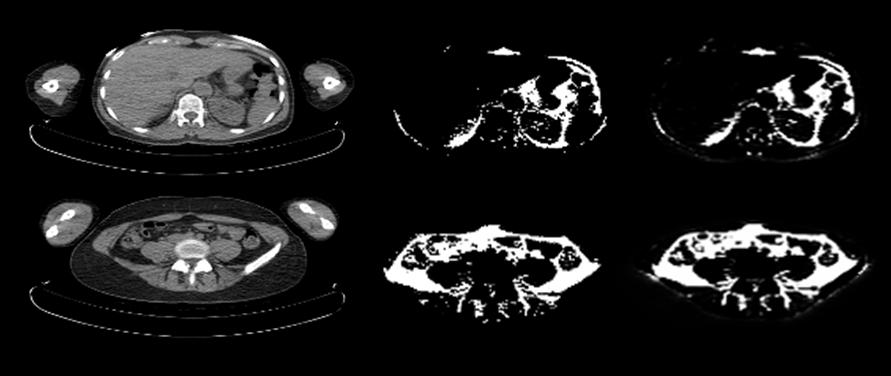}}
   \caption{\footnotesize{Examples of visceral fat segmentation: (a) the acquired CT, (b) ground truth, (c) segmented VAT using U-Net based architecture.}}
   \label{fig:fig_Unet_CT_V_Resutls}
\end{figure}

Our trained U-Net segmentation networks reveal an average dice of 97.46\% and 94.33\%, respectively in segmenting the subcutaneous and visceral fat for the test data (a-CT scans). Some visual results are shown in Fig.~\ref{fig:fig_Unet_CT_Resutls} and Fig.~\ref{fig:fig_Unet_CT_V_Resutls} for both SAT and VAT, respectively. 

\begin{figure}[!t] 
 \centering

\subfloat[\qquad\qquad\qquad\qquad(b)\qquad\qquad\qquad\qquad(c)]{\includegraphics[width = 80mm, height = 34mm]{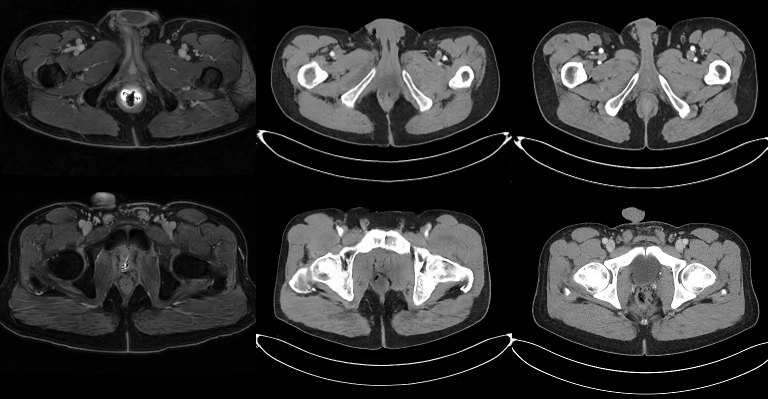}}

\caption{\footnotesize{MR and synthetic CT slices from two patients: (a) acquired MR slice, (b) acquired CT slice, and (c) generated synthetic CT slice.}}
\label{Fig_app0}
\end{figure}

We trained our C-GAN using the s-CT and MR slices in abdomen region of patients in an unpaired fashion. It must be noted that 3D images in these two modalities do not necessarily occupy the exact same physical space due to inevitable differences between scanning devices, field-of-view, and patient's physical conditions. Here, we tested our proposed framework using 320 randomly chosen MR slices as the test dataset and transformed them into their equivalent s-CT slices. \textcolor{black}{We compared our resulting s-CT with a-CT slices of these patients to quantify the performance of our algorithm (figure~\ref{Fig_app0})}. Our expert radiologist scored the generated images using the rules presented in Table \ref{tab:table0}. Our qualitative investigation showed that the s-CT scans got an average score of 4.16 out of 5. The lowest scores belonged to slices in upper lung area, where the network was unable to see enough MR samples due to the limited field-of-view in MR acquisition.

\subsection{Fat tissue segmentation in abdomen MRI}
\begin{figure}[!t] 
 \centering
\subfloat[\qquad\qquad\qquad\qquad(b)\qquad\qquad\qquad\qquad(c)]{\includegraphics[width = 80mm, height = 35mm]{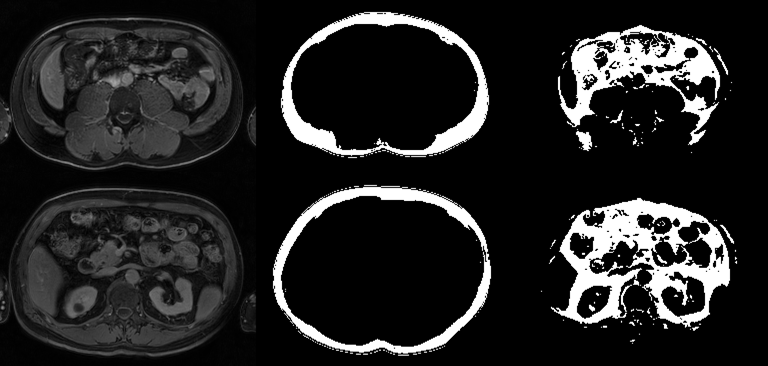}}
\caption{\footnotesize{Results for subcutaneous and visceral fat segmentation: (a) the acquired MR slice, (b) subcutaneous fat segmentation result, and (c) visceral fat segmentation result.}}
\label{Fig_app1}
\end{figure}

 We employed our trained U-Net models for visceral and subcutaneous fat segmentation in CT images to segment fat in s-CT (obtained using C-GAN) slices. The outputs of VAT and SAT segmentation in s-CT slices were assigned success scores (according to Table \ref{tab:table1}) between 0-5 by expert radiologist. The average success scores in our experiment was $3.80$ for visceral fat segmentation and $4.54$ for subcutaneous fat segmentation. Most failures in correct segmentation were observed in the lung base-upper abdomen region. In these regions, C-GAN failed to generate accurate s-CTs due to intensity overlap between visceral fat and lung parenchyma in fat-saturated T1-MRI which were used for training. A few representative examples of segmentation results are presented in Fig.~\ref{Fig_app1}.
\label{sec:pagestyle}

\section{CONCLUSIONS}
We constrained a C-GAN to perform a forward-backward 2-D transformation between MR and CT scans and used it to transform MRI slices into s-CT slices. We segmented VAT and SAT in the generated s-CT slices. The segmentation labels were assigned (with the assumption of a spatial registration between MRI and s-CT scans) to the corresponding MRI scan. We used several prepossessing techniques to improve the performance of our proposed algorithm. The results are significantly accurate as are confirmed by expert radiologist. Our proposed solution is innovative and can be extended to other bio-medical applications where ground truth labels do not exist (or difficult to obtain) for certain imaging modalities but are easier to obtain for another modality. This idea would benefit the field of medical image analysis, where labeled data is scarce. The results are also significant for clinical studies which could benefit from fat segmentation and quantification in MRI.  


\addtolength{\textheight}{-12cm}   








\bibliographystyle{IEEEbib}
\bibliography{root}
\end{document}